\documentstyle[psfig,aps,prl,amsfonts,amssymb]{revtex}

\begin{document}
\title{Bures Geometry of the Three-Level Quantum Systems. II}
\author{Paul B. Slater}
\address{ISBER, University of
California, Santa Barbara, CA 93106-2150\\
e-mail: slater@itp.ucsb.edu,
FAX: (805) 893-7995}

\date{\today}

\draft
\maketitle
\vskip -0.1cm

\begin{abstract}
For the eight-dimensional Riemannian manifold comprised by the three-level
quantum systems endowed with the Bures metric, we numerically approximate
the integrals over the manifold of several functions of the curvature and
of its (anti-)self-dual parts.
The motivation for pursuing 
this research is to elaborate upon the findings of
Dittmann in his paper, ``Yang-Mills equation and Bures metric''
(Lett. Math. Phys. 46, 281-287 [1998]).
\end{abstract}

\vspace{.2cm}
\hspace{1.5cm} Keywords: Bures metric, Yang-Mills fields, 
three-level quantum systems, Pontryagin classes,

\vspace{.005cm}

\hspace{3.3cm} action, numerical integration, octonionic instanton 
\hspace{1.5cm}

\vspace{.15cm}

\hspace{1.5cm} Mathematics Subject  Classification (2000): 81T13, 53C07, 58Z05

\pacs{PACS Numbers 03.65.-w, 02.40.Ky, 02.60.Jh}

\tableofcontents

\vspace{.1cm}
\section{Introduction}
A metric of particular interest --- the {\it minimal} member of the 
(nondenumerable) class of
monotone metrics \cite{petzsudar,les} --- that can be attached to 
the $(n^2-1)$-dimensional convex sets of  $n$-level
quantum systems is the Bures metric 
\cite{uhlmann,hubner1,hubner2,BRAUN,slatepj}
(cf. \cite{sjoqvist}). 
 Dittmann 
\cite{ditt1} established that ``the  
connection form
(gauge field) related to the generalization of the Berry phase to mixed 
states proposed by Uhlmann satisfies the source-free Yang-Mills equation
$*D*D \omega$, where the Hodge operator is taken with respect to the 
Bures metric on the space of finite-dimensional nondegenerate density
matrices'' (cf. \cite[p. 207]{tian}).
Let us also note that Bilge {\it et al} \cite{bilge,bilge2}, 
amongst others \cite{tchrakian,corrigan,grossman} (cf. 
\cite[sec. 3.f.3]{gursey}, \cite{chandia,kim}),
have studied the properties of Yang-Mills fields in {\it eight} dimensions
(in contrast to the
original, principally studied case of  {\it four} \cite{ora}).
Since the three-level ($n=3$) quantum systems do, in fact,
 form an {\it eight}-dimensional
convex set \cite{bloore} --- the $n \times n$ density matrices, in general,
forming $(n^2-1)$-dimensional convex sets --- 
it is of obvious interest to attempt to apply the analyses
of Bilge {\it et al} in light of these 
findings  of Dittmann concerning the relation between Yang-Mills 
fields and the Bures metric \cite{ditt1}.
Such an effort constitutes, by and large, the substance of this 
communication. In the 
presentation of our findings, we will conform to the notation
and conventions employed in \cite{bilge}, as detailed in the appendix
to that paper, although we should indicate that in a related paper 
\cite{bilgediff}, 
Bilge  replaces the notation used in 
\cite{bilge} for the inner 
product $(F,F)$ of an $SO(n$)-valued curvature form $F=(F_{ab}$) 
with $<F,F>$. We are compelled, however, to point out to the reader
that in our computations we have not used the $F_{ab}$'s 
themselves, but rather as ``proxies'' for them, 
the $R_{ab}$'s, describing the curvature
of the Riemannian manifold (cf. \cite[eqs. (4), (5)]{acharya} 
\cite[eq. (8)]{koz}).

Dittmann had noted \cite{ditt1} 
``that in affine coordinates (e. g. using the Pauli
matrices for $n>2$) the [Bures] 
metric becomes very complicated for $n>2$ and no 
good parameterization seems to be available for general $n$''. In 
the first part of this two-part paper 
\cite{slat1}, we attempted to fill this lacuna, by
deriving the $8 \times 8$ Bures metric tensor making use of a 
recently-developed  Euler-angle parameterization of the 
$3 \times 3$ density matrices \cite{byrd,sud}. 
Six Euler angles (denoted $\alpha,\gamma,a,\beta,b,\theta$), used in
parameterizing an element of $SU(3)$,
are employed. Additionally, two (independent) eigenvalues
 ($\lambda_{1},\lambda_{2}$) of the
$3 \times 3$ density matrices enter
into this  ``Schur-Schatten''-type
parameterization of the $3 \times 3$ density matrices \cite{twam,hase}.
(Of course, by the requirement of unit trace, 
$\lambda_{3} = 1 - \lambda_{1}- \lambda_{2}$.)
Several of the tensor elements were then, in fact, found to be identically
zero. (The $8 \times 8$ matrix tensor decomposes into
a $6 \times 6$ block and a $2 \times 2$ one, in correspondence to the six 
Euler angles and the two independent eigenvalues.) 
However, we were not able to report 
concise symbolic expressions in
\cite{slat1} for {\it all} the tensor elements.

\section{Analyses}

Subsequent to 
the issuance of Part I \cite{slat1}, we have further pursued the 
analytical matters there, 
managing  to derive
symbolic expressions 
in relatively concise form for all the entries of the $8 \times 8$ Bures metric
tensor ($g_{ij}$), as well as its inverse ($g^{ij}$). Several of the entries,
however, particularly two in the inverse, remain
rather relatively cumbersome in nature, so far resisting further 
simplification.
By way of illustration,
making use of 
the LeafCount command in MATHEMATICA to measure the complexity of an
expression, the largest leaf count (that is, the ``total number of 
indivisible subexpressions'') for any single one of 
the  $g_{ij}$'s is 352, while the largest
for any of the $g^{ij}$'s is 233, 
aside from the two relatively complicated ones, having  scores
of 921 and 1,744. 
We note here that all these entries prove to be {\it independent} of the 
Euler angle $\alpha$. Additionally, 
introducing the transformation $\gamma=
\tau-a$, all the entries also become independent of the Euler angle 
$a$. (The entry with the largest leaf count is $g^{\tau \tau}$, 
and the second largest, $g^{\tau a}$.) Therefore, the 
two tensors  are 
fully parameterizable with {\it six}, rather than eight variables.
This reduction in dimensionality is exploited 
in the numerical integrations reported below.

We noted that the two 
expressions (both symmetric in $\lambda_{1}$ and $\lambda_{2}$)
\begin{equation} \label{sub1}
A= 3 - 7 \lambda_{1} + 4 \lambda_{1}^{2} + 7 (-1 + \lambda_{1})
\lambda_{2} + 4 \lambda_{2}^2
\end{equation}
and 
\begin{equation} \label{sub2}
B= 4 \lambda_{1}^{3} + \lambda_{1}^{2} (-9 +5 \lambda_{2}) + \lambda_{1}
(-1 +\lambda_{2})(-7 +5 \lambda_{2}) + (-1+\lambda_{2}) (2 + \lambda_{2} 
(-5 + 4 \lambda_{2}))
\end{equation}
frequently occurred in the tensor elements. 
These two expressions are displayed in Figs. 1 and 2, making use of
the transformation employed in \cite{byrd}: $\lambda_{1} =
\cos^{2}{\zeta_{1}}$, $\lambda_{2} = 
\sin^{2}{\zeta_{1}} \cos^{2}{\zeta_{2}}$.
\begin{figure}
\centerline{\psfig{figure=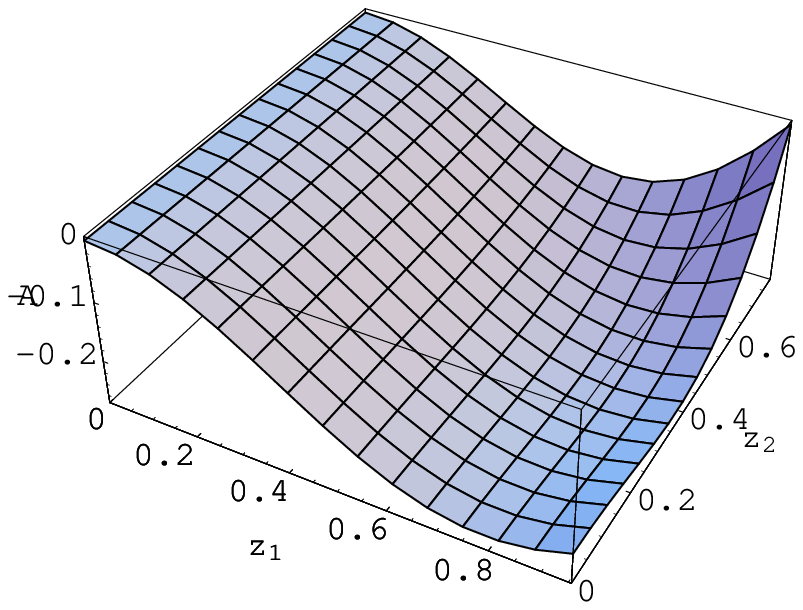}}
\caption{The subexpression $A$ --- that is (\ref{sub1}), after conversion to
spherical coordinates --- appearing in several tensor elements}
\end{figure}
\begin{figure}
\centerline{\psfig{figure=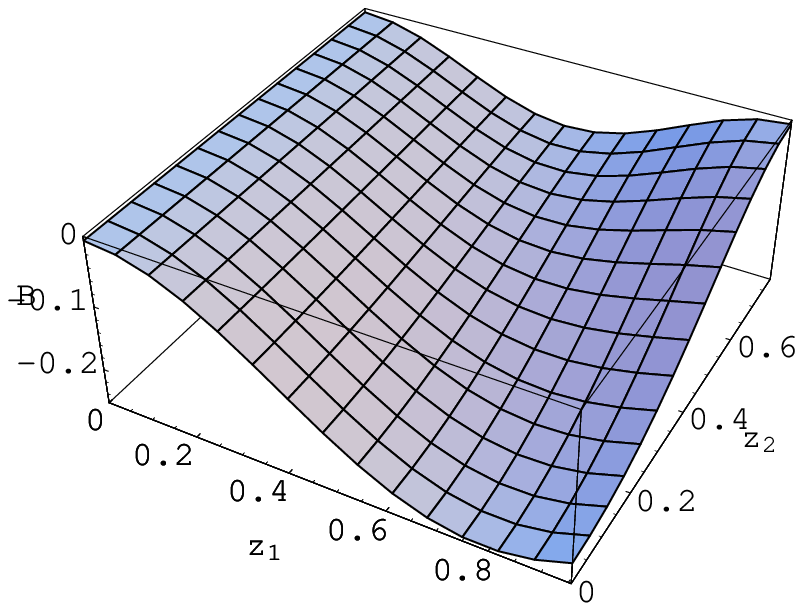}}
\caption{The subexpression $B$  --- that is (\ref{sub2}), after conversion 
to spherical coordinates --- appearing in several tensor elements}
\end{figure}
For example,
\begin{equation}
g_{\theta \theta} =- {B +A (\lambda_{1} -\lambda_{2}) \cos{2 b} 
\over 2 (-1 +\lambda_{1}) (-1 +\lambda_{2})},
\end{equation}
\begin{equation}
g^{\beta \beta} = - {(B+A (\lambda_{1} -\lambda_{2}) \cos{2 b} )
\csc^{2}{\theta}
\over 8 (-1 + 2 \lambda{_1} + \lambda_{2})^2 
(-1 + \lambda_{1} + 2 \lambda_{2})^2},
\end{equation}
\begin{equation}
g^{\alpha \alpha} = {g^{\beta \beta} \csc^{2}{\beta} \sec^{2}{\beta} \over 4}.
\end{equation}
In the two $6 \times 6$ (Euler angle) blocks of the Bures metric tensor
and its inverse, 
the identically zero elements are $g_{\tau b}, g_{\tau \theta},
g_{a b}, g_{a \theta}, g_{b \theta}, g^{\alpha \beta}, g^{a b}$ and
$g^{a \theta}$.

The volume element of the Bures metric, in particular for the case
$n=3$,
\begin{equation} \label{ve}
v. e. = \sqrt{|g|} \mbox{d} \alpha \mbox{d} a \mbox{d} \tau \mbox{d} \beta
\mbox{d} b \mbox{d} \theta \mbox{d} \lambda_{1} \mbox{d} \lambda_{2},
\end{equation}
where
\begin{equation}
\sqrt{|g|} = 
{\sin{2 b} \sin{2 \beta} \sin^{2}{\theta} \sin{2 \theta} 
\over 8  \sqrt{ \lambda_{1} \lambda_{2} \lambda_{3}}} 
{ (\lambda_{1} -\lambda_{2})^2 (\lambda_{1}-\lambda_{3})^2 (\lambda_{2}- \lambda_{3})^2 
\over (\lambda_{1}+\lambda_{2}) (\lambda_{1}+ \lambda_{3}) (\lambda_{2} + \lambda_{3})},
\end{equation}
has been determined from
certain general considerations \cite{hall,slaterhall}. 
We were 
then able, using this formula as a test,
 to numerically validate  our calculations of the $g_{ij}$'s
to a very high degree of accuracy.
Also, we were able to reproduce  --- using our 
intermediate
computation of the Ricci tensor --- the results of Dittmann \cite{ditt2}
(cf. \cite{ditton})
for the scalar curvature of the Bures metric, as applied to the $n=3$ case,
\begin{equation}
s. c. = {2 (28 e_{3} -49 e_{2} -9) \over e_{3}-e_{2}},
\end{equation}
where use is made of  the elementary symmetric polynomials,
\begin{equation}
e_{3} = \prod_{j=1,3}{\lambda_{j}}, \qquad e_{2} = \lambda_{1} -\lambda_{1}^{2}
+ \lambda_{2} - \lambda_{1} \lambda_{2} - \lambda_{2}^2.
\end{equation}
While the convex set of $n \times n$ density matrices endowed with the
Bures metric is of {\it constant} curvature for $n=2$, it is not even a {\it locally
symmetric} space for $n>2$ \cite{dittt} (cf. \cite{tanno}).
Dittmann \cite{ditt2} argues that the scalar curvature achieves its
{\it minimum}
 for the fully mixed states, while others \cite{michor}, using
another monotone metric (the {\it Kubo-Mori} one), argue that the scalar curvature
is {\it maximum} for the fully mixed states. This 
disparity is clearly a matter of
convention, depending upon what sign is 
designated for the Ricci tensor.
(Somewhat relatedly, 
Grasselli and Streater have shown that ``in finite dimensions, the only
monotone metrics on the space of invertible density matrices for which
the (+1) and (-1) affine connections are mutually dual are constant multiples
of the Kubo-Mori metric'' \cite{grasselli}.)

Additionally, it was interesting to observe that numerical evidence we
adduced
indicated that a certain necessary and sufficient condition for 
{\it Riemannian} connections to satisfy the Yang-Mills 
equation \cite{alex1,alex2,chao},
\begin{equation}
(\nabla^{g}_{X} Ric)(Y,Z) = (\nabla^{g}_{Y} Ric)(X,Z),
\end{equation}
was {\it not} fulfilled in our particular situation. 
Here $X,Y,Z$ denote arbitrary smooth vector
fields on the manifold in question --- that is, the eight-dimensional
convex set of $3 \times 3$ density matrices --- which we will denote 
by $M_{8}$, and $\nabla^{g}_{X}$ the covariant derivative with respect to $X$.

We utilized our determination of the $g_{ij}$'s and $g^{ij}$'s
in performing the same type of calculations that have been shown in 
\cite{bilge} to 
provide  upper bounds on Pontryagin
numbers for $SO(n)$ 
Yang-Mills fields. These bounds are 
\begin{equation} \label{bound1}
\int_{M_{8}} (F,F)^2 \geq k \int_{M_{8}} p_{1}(E)^2,
\end{equation}
where
\begin{equation} 
(F,F) =2 \sum_{a=1}^{8} \sum_{b > a}^{8} (F_{ab},F_{ab}),
\end{equation}
and
\begin{equation} \label{bound2}
\int_{M_{8}} (F^2,F^2) \geq k' \int_{M_{8}} p_{2}(E).
\end{equation}
Here, the field strengths
$F_{ab}$'s  denote the 
skew-symmetric coordinate components of $F$, an $SO(n)$-valued 
curvature two-form on the Euclidean
space ${{\mathbb{R}}}^{{{8}}}$.
For the $ij$-entry of an  $8 \times 8$ matrix $F_{ab}$ 
was computed by us as
$R_{ijab}$ --- that is, an 
element of the fully covariant (0,4)-Riemann 
curvature tensor \cite[eq. (6.6.1)]{weinberg} 
(cf. \cite[eqs. (4), (5)]{acharya}). Also, 
$k$ and $k'$ are certain constants, while  
$p_{1}(E)$ and $p_{2}(E)$ are the first and second Pontryagin classes
for the 
bundle $E$ in question. The principal $U(n)$-bundle $E$ ($n=3$) utilized 
by Dittmann \cite{ditt1}  is
 $Gl(n,\mathbb{C}) \rightarrow $ $ 
{Gl(n,\mathbb{C})}/{U(n)}$, 
where $Gl(n,\mathbb{C})$ denotes the general linear group over 
the complex numbers $\mathbb{C}$, that is 
the nonsingular $n \times n$ matrices 
with complex entries \cite[ex. 5.1]{kn}. (The Pontryagin number for a {\it four}-dimensional
 compact manifold $M_{4}$ is equal to the integral over $M_{4}$ of a 
representative of the first, and unique 
Pontryagin class of the bundle, that is the 
Chern number $\int_{M_{4}} \gamma_{2}$ of the 
complexified bundle \cite{choquet}.)

The first of the two bounds (\ref{bound1})
 was reported in \cite{bilge} 
and the second (\ref{bound2}) 
in \cite{grossman}. The first is achieved by {\it strongly}
[anti-] self-dual Yang-Mills fields, while the second is too restrictive 
in this regard \cite{bilge}. An additional curvature invariant
is $(\mbox{tr}F^2,\mbox{tr}F^2)$, so a ``generic action density'' 
can be taken to be  of the form \cite[eq. (17)]{bilge}
\begin{equation} \label{lfh}
 p (F,F)^2 + q (F^2,F^2) + r (\mbox{tr} F^2,\mbox{tr} F^2).
\end{equation}
(``The action density should be written in terms of the local curvature
2-form matrix in a way independent of the local trivialization of the
bundle. Hence it should involve invariant polynomials of the local
curvature matrix. We want to express the action as an inner product
in the space of $k$-forms, which gives a quantity independent of the
local coordinates'' \cite{bilge}.)
\subsection{Monte-Carlo numerical integrations}
To evaluate the integrals over $M_{8}$ of the three 
curvature invariants occurring in the generic action (\ref{lfh}), we   
pursued a Monte-Carlo strategy \cite{halton,james,krommer}
in which we {\it randomly}
 selected points in the six-dimensional hyperrectangular
domain determined by the six active  parameters (four  Euler 
angles --- $\tau, \beta, b, \theta$ --- and 
two independent eigenvalues reexpressed using
spherical coordinates, that is, $\lambda_{1} = \cos^{2}{\zeta_{1}}$,
$\lambda_{2} = \sin^{2}{\zeta_{1}} \cos^{2}{\zeta_{2}}$ \cite{byrd}).
This six-dimensional domain 
has one side of length $\pi$ (corresponding to the range 
of $\tau$), three sides (for $\beta, b$ and $\theta$) of length
$\pi /2$, one side
(for $\zeta_{1}$) of length $\cos^{-1}{\pi/3} \approx .955317$ and  one 
 (for $\zeta_{2}$) of length $\pi /4$.
\cite{byrd}. The two remaining (seventh and eighth) 
sides --- associated with the Euler 
angles $\alpha$ and $a$, which we have noted 
are absent from the simplified transformed expressions for the 
$g_{ij}$'s \cite{slat1} --- are of length
$\pi$.  
At each of the randomly selected points,
we computed  the products of the volume element (\ref{ve}) 
with the quantities $(F,F)^2$, $(F^2,F^2)$,
$(\mbox{tr} F^2,\mbox{tr} F^2)$ --- as well as $(F^3,F^3)^{2/3}$ 
and $(F^4,F^4)^{1/2}$. 
(In these MATHEMATICA 
computations, we utilized the relation $\mbox{det}(I +t F) 
=\sum_{k=0}^n \sigma_{k} t^k$, where $\sigma_{k}$ are invariants of
the local trivialization \cite[eq. (12)]{bilge}, \cite{bilgeJMP}.) 
We multiplied the averages 
of each of the five  products over the randomly chosen points 
by the full eight-dimensional Euclidean hyperrectangular volume, which is
${\pi^7 \over 32} \cos^{-1}{1 \over \sqrt{3}} \approx 90.1668$.
For 21,238 randomly generated points, our 
set of five results (exactly as computed, without any ``rounding'') 
is presented in the {\it first} line of Table I.

\begin{table} 
\begin{tabular}{r||r|r|r||r|r||}
field & $\int_{M_{8}}(F,F)^2$ & $\int_{M_{8}}(F^2,F^2)$ & 
 $\int_{M_{8}}(\mbox{tr} F^2,\mbox{tr} F^2) $ &
$\int_{M_{8}} (F^3,F^3)^{2/3} $ &  $ \int_{M_{8}}(F^4,F^4)^{1/2}$ \\
\hline \hline
Bures metric & .00174878  & $ 9.91872   \cdot 10^{-14}$ 
& .00174878  & $  7.90173  
\cdot 10^{-22} $  &  $ 5.07518   \cdot 10^{-29} $ \\
anti-self-dual part & 2.0692  & 1.5519  & 3.62111  & 1.41  & 1.09736  \\
self-dual part & 18.6228   & 7.75951  & 26.3823  &  5.15961 & 3.29208  \\
\end{tabular}
\label{pop}
\caption{Curvature invariants approximated by evaluating the 
integrands at  21,238 
points randomly selected in the six-dimensional hyperrectangle of active 
parameters}
\end{table}
We also, following 
Baulieu and Shatashvili \cite[eqs. (3.4)-(3.6)]{baulieu} 
(cf. \cite[p. 207]{tian}, \cite[eq. (2.9)]{gao}),
decomposed the curvature two-form for the $8 \times 8$ 
Bures metric tensor into two $Spin(7)$-irreducible
components, 
according to $ 28 =21 \otimes 7$, which can be called {\it 
self-dual}
and {\it anti-self-dual} respectively. We then evaluated the same
curvature norms. The corresponding Monte-Carlo results for the {\it same} set
of 21,238 randomly selected points are presented in the {\it second}
 and {\it third} lines
of Table I. It clearly appears, then, that the Yang-Mills field for the
Bures metric on the three-level quantum systems is neither ``self-dual''
nor ``anti-self-dual'' (cf. \cite{sibner,wong,sadun,bor,burz}).

Since all the five integrands used in Table I are nonnegative in nature,
it appears that for the Yang-Mills/Bures field (corresponding to the 
first line of the table), $(F^3,F^3)$ 
and $(F^4,F^4)$, at least, are themselves zero.
Also, we see that in in this same first line
(but not the second and third), the integral of
$(F,F)^2 ${\it equals} that of $(\mbox{tr} F^2,\mbox{tr} F^2)$.
Here \cite[eq. (21)]{bilge}
\begin{equation}
\mbox{tr} F^2 = -2 \sum_{a=1}^{8} \sum_{b>a}^{8} F_{ab}^{2}.
\end{equation}
(For the octonionic $SO(7)$ instanton  two-form $F$, the 
 four-form $\mbox{Tr} F \wedge F  $ is neither self-dual nor anti-self-dual 
\cite{duff}.) 
The $8 \times 8$ skew-symmetric matrices $F_{ab}$ must have 
four pairs of  imaginary
eigenvalues of opposite sign. 
(``Strong self-duality or strong anti-self-duality can be characterized 
by requiring the equality of the absolute values of the 
[eight] eigenvalues'' \cite{bilge}.) For the 
eight-dimensional Yang-Mills field $F$ over the three-level quantum
systems, our analyses indicate that one of these four pairs is always
(0,0) --- clearly indicative of some form of degeneracy. 
(Riemannian manifolds the skew-symmetric curvature operators of which
have {\it constant} eigenvalues have been the subject of study \cite{gilkey},
including the case of {\it eight}-dimensional manifolds \cite{gilkey2}.)
This ensures that the 
determinant (the product of the eigenvalues), as well as the
$(F_{ab}^4,F_{ab}^4)$'s themselves,  and thus $(F^4,F^4)$, are zero.
The apparent zero nature of $(F^2,F^2)$ and $(F^3,F^3)$
can not be directly explained in this manner, that is, by the zero nature
of {\it one} pair of eigenvalues.
(Let us point out here the possible relevance for our study, yielding 
a ``small action'',  of
\cite{yunmei}.)
\subsection{Numerical integrations using regular lattices}
As somewhat of a ``cross-check'' on  the  Monte-Carlo analyses 
reported above in Table I 
(motivated, in part, by some apparent instabilities in early, preliminary
analyses), we also undertook  numerical
integrations by evaluating the various 
curvature quantities not at randomly selected
points in the six-dimensional hyperrectangle, but at the nodes of regular
lattices superimposed on this hyperrectangle. For a $2^6=  64$-point lattice
(dividing each 
parameter range into two equal segments and taking the midpoints of
each segment as coordinates), we found 

\begin{table} \label{hus}
\begin{tabular}{r||r|r|r||r|r||}
field & $\int_{M_{8}}(F,F)^2$ & $\int_{M_{8}}(F^2,F^2)$  
& $\int_{M_{8}}(\mbox{tr} F^2,\mbox{tr} F^2)$ &
$\int_{M_{8}} (F^3,F^3)^{2/3} $ &  $\int_{M_{8}}(F^4,F^4)^{1/2}$ \\
\hline \hline
Bures metric & .00109779 & $1.399 \cdot 10^{-26} $ & .00109779 & 
$4.297 \cdot 10^{-36}$ &  $7.231 \cdot 10^{-44} $ \\
anti-self-dual part & 1.77679  & 1.3326  & 3.10939 & 1.21074  & .942287  \\
self-dual part & 15.9911 & 6.66298  & 22.6541 & 4.43048  & 2.82686  \\
\hline
SD part minus ASD part & 28.4287 & 21.3215 & 49.7502  & 19.3719 & 15.0766 \\
\hline
\end{tabular}
\caption{Curvature invariants approximated by evaluating 
the integrands at the nodes
of a 64-point lattice imposed upon the six-dimensional hyperrectangle 
of active parameters}
\end{table}
A {\it fourth} line is included in Table II 
 for the {\it difference} \cite[eq. (3.4)]{baulieu}
\begin{equation}
 ^{\dagger} F = F^{+} - F^{-},
\end{equation}
of the self-dual $(F^{+})$ and anti-self-dual ($F^{-}$) fields,
so
\begin{equation}
F^{\pm} = {1 \over 2} (F \pm ^{\dagger} F).
\end{equation}

For a (finer) $3^6 = 729$-point lattice, we obtained
\begin{table}
\begin{tabular}{r||r|r|r||r|r||}
field & $\int_{M_{8}}(F,F)^2 $ & $\int_{M_{8}}(F^2,F^2) $ & 
$\int_{M_{8}} (\mbox{tr} F^2,\mbox{tr} F^2)$ & $ \int_{M_{8}}(F^3,F^3)^{2/3}$
& $\int_{m_{8}} (F^4,F^4)^{1/2}$ \\
\hline \hline
Bures metric & .00029131  & $1.22054 \cdot 10^{-26}$  &  .00029131 & $
 1.15258 \cdot 10^{-35}$ & $ 1.17924 \cdot 10^{-42} $ \\
anti-self-dual part & 2.11051  & 1.58288  & 3.69339  & 1.43814  & 1.11927  \\
self-dual part & 18.9946  & 7.9144  & 26.909  & 5.2626 & 3.3578  \\
\end{tabular}
\caption{Curvature invariants approximated by evaluating the 
integrands at the nodes
of a 729-point lattice imposed upon the six-dimensional hyperrectangle of 
active parameters}
\end{table}
Additionally, for a $4^6 = 4096$-point lattice, we computed
\begin{table}
\begin{tabular}{r||r|r|r||r|r||}
field & $\int_{M_{8}}(F,F)^2$ & $\int_{M_{8}}(F^2,F^2)$  & $ \int_{M_{8}}
(\mbox{tr} F^2,\mbox{tr} F^2)$ & $ \int_{M_{8}} (F^3,F^3)^{2/3}$  &
$\int_{M_{8}} (F^4,F^4)^{1/2}$ \\
\hline \hline 
Bures metric & .000824594 & $5.54578 \cdot 10^{-26} $ & 
.000824594 & $ 3.95448 \cdot 10^{-36} $ & $ 9.09798 \cdot 10^{-43} $ \\
anti-self-dual part & 2.08472 &  1.56354 & 3.64827 &  1.42057 &  1.10559 \\
self-dual part & 18.7625 & 7.81771 &  26.5802 & 5.19831 &  3.31677 \\
\end{tabular}
\caption{Curvature invariants approximated by evaluating the 
integrands at the nodes of
a 4096-point lattice imposed upon the six-dimensional hyperrectangle of 
active parameters}
\end{table}

We see that the results here, in their overall aspects, are supportive of
our much  
more extensive (and thus we are inclined to believe more accurate) 
Monte-Carlo analysis in Table I --- to which we are in the
process of adding still more randomly selected points.
(It might be noted that none of our
tabulated results --- for the anti-self-dual and self-dual 
components --- when divided by $\pi^{n}$ , for $n=4$ or other small
integers,
gives indications of being integral in nature
\cite[eq. (7)]{chandia2}.
Also, there might seem to be a question of whether our tabulated
results should be ``corrected'' by a factor of $3! = 6$, 
that is the number of ways
of permuting the rows and columns of a $3 \times 3$ density matrix,
but doing so would lead to density matrices that are simply permutations of
one another being represented by {\it distinct} points in the manifold 
\cite{byrd},
\cite[sec. III.B]{slaterhall}.)

\subsection{Computation of Yang-Mills action functionals} \label{gju}

For a $G$-bundle $E->X$, with connections $A$, the Yang-Mills functional
can be expressed in the eight-dimensional case as \cite[eq. (2.4)]{gao}
\begin{equation}
S_{YM}[A] = {1 \over {2 g^2}} \int_{X} d^{8} x \sqrt{g} \mbox{Tr} 
F_{\mu \nu} F^{\mu \nu} \equiv {1 \over {2 g^2}}  {|| F ||}^{2},
\end{equation}
where ${1 \over g^2}$ represents the coupling constant of the gauge fields.
For the Yang-Mills gauge field defining the Bures metric 
\cite{ditt1}, based on 14,151
randomly chosen points in the six-dimensional hyperrectangle of active
parameters, we obtained a value of 
.0145485   for ${|| F ||}^{2}$. For
$ {||  F^{-} || }^{2}$,
the comparable value was much larger, that is 
$5.33255 \cdot 10^6$, and almost identical  but  slightly larger, 
that is $5.33268 \cdot 10^6$, for ${||  F^{+} ||}^{2}$.
``It can be shown that any self-dual 2-form defined by the above criteria
[for self-duality in dimensions greater than 4] satisfies the 
Yang-Mills equations. However, the corresponding Yang-Mills action
need not be extremal'' \cite[p. 4804]{bilgeJMP}. ``\ldots it is no longer
true that a configuration satisfying our duality equation 
[in arbitrary even dimensions] automatically
corresponds to a minimum of the [Yang-Mills] action'' \cite[sec. 2.4]{bais}.

Also Gao and Tian assert \cite[eq. (2.5)]{gao}
\begin{equation} \label{gt25}
S_{YM}[A] = {1 \over g^2 (B +\sqrt{B^2 +8 A})} 
\lgroup 2 \int_{X} \Omega \wedge \mbox{Tr} (F \wedge F) 
+ \sqrt{B^2 +8 A} {||F^{+}||}^{2} \rgroup,
\end{equation}
where \cite[eq. (2.8)]{gao} $A=6, B= -4$ and the four-form $\Omega$
is a ``bispinor'' \cite[eq. (2.7)]{gao}. They state that
``the Yang-Mills action can be written as a non-negative term
proportional to ${||F^{+}||}^{2}$, plus a topological invariant.
Clearly, such an action will reach its minimal values at $F^{+} = 0$''
\cite[p. 3]{gao} (cf. \cite{sibner,wong,sadun,bor,burz}).

In an exchange with Y. H. Gao on this matter, he has written:
``Tentatively, the puzzle you mentioned (i. e. the Yang-Mills functional
for the (anti) self-dual field is larger than for the full field) could stem
from the problem of constructing a topological term using numerical methods.
In the analytic world the difference between $||F||^{2}$ and 
$4 ||F_{+}||^{2}$ is 'topological' in the sense that its variation with 
respect to the connection vanishes. So, locally a minimum of 
$||F_{+}||^{2}$ is also a minimum of the original Yang-Mills action, and
$||F||^{2}$ should be larger than $||F_{-}||^{2}$ provided $F$ and $F_{-}$
belong to the same 'topological class' (namely the topological term has the
same value). In numerical computations, however, manifestation of
topological invariance for an integral might not be as easy as in the analytic
world. It could be the case that when $F$ is changed into $F_{-}$, the value
of the ``topological term'' is also changed. In that case the argument 
in the analytic consideration would be no longer valid. Thus a naive
suggestion is that during comparison between $||F||^{2}$ and 
$||F_{-}||^{2}$ numerically, one should also monitor possible changes
in the topological term, to rule out the possibility that $F$ and $F_{-}$
are not really in the same topological class.''

So, it appears that it would be of interest in this matter to, in addition
to our other computations, attempt to numerically approximate the term
\begin{equation}
\int_{X} \Omega \wedge \mbox{Tr} (F \wedge F),
\end{equation}
in (\ref{gt25}) by itself, as well as with $F$ replaced in it 
 by $F_{+}$ as well
as $F_{-}$.

\subsection{Monopole equations with $Spin(7)$-holonomy}
Motivated by the Seiberg-Witten equations \cite{sw}, Bilge {\it et al} formulated eight-dimensional monopole equations as
follows 
\cite[eq. (24)]{bilgemonopole}:
\begin{equation} \label{bdk}
D_{A}(\Psi) =0, \qquad {\rho}^{+}(F_{A}^{+}) = (\Psi {\Psi}^{*})^{+}.
\end{equation}
Here (in the notation of \cite{bilgemonopole}) 
$(\Psi {\Psi}^{*})^{+}$ is the orthogonal projection of $\Psi {\Psi}^{*}$ 
onto the spinor subbundle spanned by ${\rho}^{+}(f_{i}), i = 2 \ldots 7$.
We numerically solved the second set of equations (\ref{bdk}) for the 
components of the spinor \cite[eqs. (3a.57), (3f.102), (3f.103)]{gursey}
(cf. \cite[p. 30]{bilgemonopole})
\begin{equation}
\Psi={1 \over 2} (e_{1} + e_{4} \mbox{i}, e_{2} +e_{5} \mbox{i},
e_{3} + e_{6} \mbox{i} , e_{0} + e_{7} \mbox{i},e_{1} - e_{4} \mbox{i},
e_{2} - e_{5} \mbox{i},e_{3} -e_{6} \mbox{i} ,e_{0} - e_{7} \mbox{i} ),
\end{equation}. Only for what we have termed the ``anti-self-dual''
component ($F^{-}$) did we obtain solutions, which always came in one of two
forms.
The first was 
$e_{2} = {u \mbox{i}  \over e_{4}}$, with $u$ equal to a {\it real} 
constant,
and the other six $e$'s all 
 set to zero, while the
second solution was 
$e_{0} = e_{3} =e_{6} =  e_{7} = 0$, 
\begin{equation}
e_{1} = {u  \mbox{i} e_{5} \over e_{4}^2 + e_{5}^{2}}, \quad
e_{2} = {u \mbox{i} e_{4} \over e_{4}^{2} + e_{5}^{2}},
\end{equation}
where $u$ is the same as in the first solution.

It remains for us to attempt to solve the first of the two sets of equations
in (\ref{bdk}), that is the Dirac equation, together with the Coulomb gauge 
condition.

\section{Discussion}

What we have considered to be the  ``anti-self-dual''
 part ($F^{-}$) of the Yang-Mills field over
the eight-dimensional convex set of three-level 
quantum systems satisfies the ``set $ b$'' of 
twenty-one equations
in \cite{bilge2}, 
\begin{equation} \label{kee}
F_{12} -F_{34} =0; \quad F_{12} -F_{56} = 0; \quad F_{12} - F_{78} =0; \quad
F_{13} + F_{24} =0; \quad F_{13} -F_{57} =0; \quad F_{13} +F_{68} = 0 \quad
\end{equation}
\vspace{-.5cm}
\begin{displaymath}
F_{14} - F_{23} =0; \quad F_{14} +F_{67} =0; \quad F_{14} +F_{58} = 0; \quad
F_{15} +F_{26} =0; \quad F_{15} +F_{37} =0; \quad F_{15} -F_{48} = 0 \quad
\end{displaymath}
\vspace{-.5cm}
\begin{displaymath}
F_{16} -F_{25} =0; \quad F_{16} -F_{38} =0; \quad F_{16} - F_{47} =0; \quad
F_{17} +F_{28} =0; \quad F_{17} - F_{35} = 0; \quad F_{17} + F_{46} =0 \quad
\end{displaymath}
\vspace{-.5cm}
\begin{displaymath}
F_{18} - F_{27} = 0; \quad F_{18} +F_{36} =0; \quad F_{18} +F_{45} =0
\end{displaymath}
that is, to the {\it negative} eigenvalue, -3, of a fourth rank tensor 
invariant under $SO(7)$. (A skew-symmetric matrix satisfying the set
(\ref{kee}) must have all its [imaginary] eigenvalues equal in absolute value 
\cite{bilge}.)

The ``self-dual'' part ($F^{+}$) of the Yang-Mills/Bures field 
satisfies the ``set $a$'' 
of seven equations \cite{bilge2},
\begin{equation} \label{keee}
F_{12} +F_{34} +F_{56} +F_{78} = 0; \quad F_{13} - F_{24} +F_{57} - F_{68} = 0;
\quad F_{14} +F_{23} -F_{67} -F_{58} =0; \quad F_{15} - F_{26} -F_{37} +F_{48} =0
\end{equation}
\vspace{-.5cm}
\begin{displaymath}
 F_{16} + F_{25} +F_{38} +F_{47} =0; \quad F_{17} -F_{28} +F_{35} 
- F_{46} = 0; \quad F_{18} +F_{27} -F_{36} - F_{45} = 0
\end{displaymath}
 that is to the
{\it positive} eigenvalue 1.

``The solutions of set $a$ and set $b$ can be viewed as analogues of
self-dual 2-forms in four dimensions from different aspects. The strongly
self-dual 2-forms, hence the solutions of set $b$ saturate various topological 
lower bounds \ldots, but they form an overdetermined system \ldots
we show that the solutions of set $b$ for an $N$-dimensional gauge group,
depend exactly on $N$ arbitrary constants, provided that the system is
consistent. Thus the set $b$ lacks the rich structure of the self-duality 
equations in four dimensions. On the other hand, the solutions of set $a$ do
not saturate the topological bounds \ldots but these equations form an 
elliptic system under the Coulomb gauge condition'' \cite{bilge2}.

``In fact, one often finds in the literature that 
[our] equations (\ref{keee})
and (\ref{kee}) are referred to as the self-duality and anti-self-duality 
equations respectively. This nomenclature suggests a symmetry between these
equations which is not present in the octonionic case since, for example, the
spaces have different dimension. In our opinion, self-duality and
anti-self-duality correspond to which way the division algebra 
$\mathbb{A}$ acts: 
if on the left or on the right, and are hence related by a change of
orientation on the manifold. Although there has been some work in the 
literature concerning equation (\ref{kee}), we believe this equation is not
as fundamental as (\ref{keee}). This can also be seen not just in the results
of the present paper but also, for example in \cite{new}, where it is 
shown that supersymmetry singles out equation (\ref{keee})'' \cite{figueroa}.
(In our examination of various papers, it even appears that what one author
calls the ``self-dual'' part, another author may call the ``anti-self-dual'' 
part --- for example, \cite{tian,gao}.)

Of obvious research interest would be companion studies to that 
(eight-dimensional one) here
of the Bures metric on various {\it four}-dimensional convex sets of
density matrices --- such as those examined in \cite{slatepj,slatqfb}.
In line with this we have 
sought to calculate the Yang-Mills functionals, as in sec.~\ref{gju}, 
for the {\it four}-dimensional convex set of $3 \times 3$
density matrices studied in \cite{slatqfb} endowed with the Bures metric
\begin{equation}
\rho={1 \over 2} \pmatrix{v + z & 0 & x -\mbox{i} y \cr
0 & 2 - 2 v & 0 \cr
x +\mbox{i} y & 0 &  v - z \cr}
\end{equation}
The Yang-Mills functionals for the original and self-dual and anti-self-dual
fields turn out to be infinite in this case, though.

In the context of $D$-dimensional Euclidean gravity, Acharya and O'Loughlin
have defined a generalization of the self-dual Yang-Mills equations as
duality conditions on the curvature two-form $R$ of a Riemannian manifold
\cite{acharya}. For $D=8$, solutions to these self-duality equations
are provided by manifolds of $Spin(7)$ holonomy \cite{joyce1,joyce2,joyce3}, 
which are necessarily Ricci-flat. 
The holonomy group is
then 21-dimensional.
Acharya and O'Loughlin replaced the field strengths in the self-dual
Yang-Mills equation
\begin{equation}
F_{\mu \nu} = {1 \over 2} \phi_{\mu \nu \lambda \rho} F_{\lambda \rho}
\end{equation}
by the components $R_{ab}$ 
 of the curvature two-form (as we have above).

Using the technique
of embedding the spin connection in the gauge connection
(cf. \cite[sec. 10.5]{kaku}), they construct
a self-dual gauge field directly from the self-dual metric. 
They let ``$G_{ab}$ be the generators of one of $SU(2), SU(3), G_{2}$ or
$Spin(7)$. The ansatz for the gauge field is $A= \gamma G_{ab} \omega_{ab}$. 
The  form index of $A$ comes from the form index of $\omega$ while the Lie 
algebra structure of $A$ comes from that of $G$. One sees then that
\begin{equation}
F = dA +A \wedge A
\end{equation}
\begin{equation}
 = \gamma G_{ab} d \omega_{ab} + \kappa \gamma^{2} G_{ab} \omega_{ac} 
\wedge \omega_{cb}
\end{equation}
where $\kappa$ is a constant that depends upon the group generated 
by $G_{ab}$. In each case it is trivial to solve for $\gamma$ giving 
$F= \gamma G_{ab} R_{ab}$. Duality of
$F$ follows from that of $R$ and the symmetry of 
$R_{\mu \nu \lambda \rho}$ between the first pair and 
second pair of indices'' \cite{acharya}.
We shall attempt to follow this prescription in our continuing research
on the topic of this paper.

The duality operator, $\phi_{a b c d}$ for $D=8$, 
is the unique $Spin(7)$-invariant
four-index antisymmetric tensor which is Hodge self-dual.
Proposition 10.6.7 of \cite{joyce3} relates the first Pontryagin class
of a manifold $M$ with a Spin-(7) structure $(\Omega,g)$ to the integral
over $M$ of $|R|^2$, where $R$ is the Riemann curvature of $g$.
Also, if $(M,\Omega,g)$ is a compact Spin(7)-mainfold, then $g$ has holonomy
Spin(7) if and only if $M$ is simply connected and the Betti numbers of
$M$ satisfy $b^3 +b_{+}^{4} = b^2 + 2b_{-}^{4} + 25$ 
\cite[Thm. 10.6.8]{joyce3}.

It has been known for some time that the Bures metric on the {\it two}-level
quantum systems is isometric to the standard metric on the three-sphere
\cite{hubner1,braun}. 
``The Bures metric for a two-dimensional system corresponds to the surface
of a unit four-ball, i. e., to the maximally symmetric three-dimensional 
space of positive curvature (and may be recognized as the spatial part of the 
Robertson-Walker metric in general relativity). This space is homogeneous and
isotropic, and hence the Bures metric does not distinguish a preferred 
location or direction in the space of density operators. Indeed, as well as 
rotational symmetry in Bloch co-ordinates (corresponding to 
unitary invariance), the metric has a further set of symmetries generated
by the infinitesimal transformations
\begin{equation}
\mathbf{r} \rightarrow \mathbf{r} + \epsilon \mathit{(1-r^2)^{1/2}} \mathbf{r}
\end{equation}
(where $\mathbf{r}$ is an arbitrary three-vector)'' \cite[p. 128]{hall}, 
$r$ being the radial distance --- the length of $\mathbf{r}$ --- in the 
Bloch sphere of two-level
systems (from the fully mixed state, $r=0$). 
Petz and S\'udar observe that ``in the case of the [Bures] metric,
the tangential component is independent of $r$'' \cite[p. 2667]{petzsudar}.

Contrastingly, Dittmann \cite{dittt} 
established that the Bures metric on the {\it three}-level
quantum systems is not a space of constant curvature nor
even locally symmetric.
(Neither, is the Einstein-Yang-Mills equation fulfilled with
a certain cosmological constant \cite{ditt1}, as it is for the
two-level quantum systems \cite{rudolph}.)
Nevertheless, we have obtained in 
Tables I-IV, strong indications that the curvature 
(known to satisfy the Yang-Mills equation \cite{ditt1}) of the Bures 
metric on the three-level quantum systems is quite flat in character.
(``quantum information manifolds are equipped with two natural flat 
connections: the mixture connection, obtained from the linear structure
of trace class operators themselves, and the exponential connection,
obtained when combinations of states are performed by adding their 
logarithms...the Bogoliubov-Kubo-Mori metric is, up to a factor,
the unique monotone Riemannian metric with respect to which the exponential
and mixture connections are dual \cite{grasselli}.)

Due to the relative complexity (high leaf count) of our current formulas
for the entries of the $8 \times 8$ Bures metric tensor and its inverse,
we have had to have recourse here, by and large, 
to {\it numerical} calculations in order to gain insights into the nature
of the associated Yang-Mills field. If further progress in simplifying
these entries can be achieved, it may be possible to proceed further
with more highly desirable exact symbolic computations.

\acknowledgments

I would like to express appreciation to the Institute for Theoretical
Physics for computational support in this research, as well as  to A. Uhlmann,
A. Bilge, R. Montgomery, D. V. Alekseevskij, 
Yi-hong Gao, D. Eardley  and D. Holz for
various helpful communications.

\end{document}